# Cascade of phase transitions and large magnetic anisotropy in a triangle-kagome-triangle trilayer antiferromagnet

*Chao Liu,[1] Tieyan Chang,[2] Shilei Wang,[1] Shun Zhou,[1] Xiaoli Wang,[1] Chuanyan Fan,[1] Lu Han,[1] Feiyu Li,[1] Huifen Ren,[3] Shanpeng Wang,[1] Yu-Sheng Chen[2] and Junjie Zhang[1]\**

[1]State Key Laboratory of Crystal Materials and Institute of Crystal Materials, Shandong University, Jinan 250100, Shandong, China

[2]NSF's ChemMatCARS, The University of Chicago, Lemont, IL 60439, United States

[3]Synergetic Extreme Condition User Facility (SECUF), Institute of Physics CAS, Beijing 101400, China

**ABSTRACT**

Spins in strongly frustrated systems are of intense interest due to the emergence of intriguing quantum states including superconductivity and quantum spin liquid. Herein we report the discovery of cascade of phase transitions and large magnetic anisotropy in the averievite $CsClCu_5P_2O_{10}$ single crystals. Under zero field, $CsClCu_5P_2O_{10}$ undergoes a first-order structural transition at around 225 K from high temperature centrosymmetric $P\bar{3}m1$ to low temperature noncentrosymmetric $P321$, followed by an AFM transition at 13.6 K, another structural transition centering at ~3 K, and another AFM transition at ~2.18 K. Based upon magnetic susceptibility and magnetization data with magnetic fields perpendicular to the *ab* plane, a phase diagram, consisting of a paramagnetic state, two AFM states and four field-induced states including two magnetization plateaus, has been constructed. Our findings demonstrate that the quasi-2D $CsClCu_5P_2O_{10}$ exhibits rich structural and metamagnetic transitions and the averievite family is a fertile platform for exploring novel quantum states.

## 1. Introduction

Understanding the origin of high-temperature superconductivity remains a big challenge and is among the current research frontiers.[1-3] Nearly forty years ago, Anderson proposed that the preexisting magnetic singlet pairs of quantum spin liquid (QSL), which is an insulating magnetic state favored by low spin, low dimensionality and magnetic frustration,[4] could become charged superconducting pairs when it is doped sufficiently strongly.[5] Anderson's theory has not been tested up to now.[6-9] The main obstacle is that no existing materials have been confirmed to be QSL due to the lack of smoking-gun evidence;[10] only a few materials are proximate QSL candidates, including $k$-(ET)$_2$Cu$_2$(CN)$_3$, herbertsmithite ZnCu$_3$(OH)$_6$Cl$_2$, and α-RuCl$_3$.[7,11,12] Herbertsmithite, which does not show any long-range magnetic order or spin glass behavior down to 20 mK,[13-15] is probably one of the most studied QSL candidates. Theoretically, electron doping with Ga on the Zn site in herbertsmithite was predicted to produce novel states including *f*-wave superconductivity and a correlated Dirac metal,[16] and hole doping with Li or Na on the Zn site was predicted to host fractional quantum Hall effect.[17] Experimentally, attempts to dope electrons/holes by substitution via various techniques in this material ended in unsuccess, one reason is that the removal of Zn from the three-dimensional (3D) crystal structure leads to decomposition.[15,18] One notable progress was the report of lithium intercalation in herbertsmithite; however, this did not lead to metallization, not to mention superconductivity.[19] Thus, it is desirable to find other QSL candidates that maintain the key ingredients of herbertsmithite (kagome lattice with Cu$^{2+}$ $S = 1/2$) but do not have its issues.

Geometrically frustrated systems such as triangular, honeycomb, kagome, hyperkagome and pyrochlore are fertile playgrounds for the exploration of novel quantum phases including quantum spin liquid.[8] Averievite, an oxide mineral represented by the formula (MX)$_n$Cu$_5$T$_2$O$_{10}$ (M = K, Rb, Cs, Cu; X = Cl, Br, I; n =1; T = P, V)[20-23] that contains trangle-kagome-triangle trilayers of Cu$^{2+}$ ($S = 1/2$), is a geometrically frustrated system and has been proposed to host QSL.[24] More importantly, unlike the 3D network in herbertsmithite,[25] averievite exhibits a quasi-2D crystal structure consisting of triangle-kagome-triangle trilayers of Cu and MO$_2$ layers between them, potentially overcoming the doping issues encountered in herbertsmithite. CsClCu$_5$V$_2$O$_{10}$ polycrystalline powders were reported to crystallize in the $P\bar{3}m1$ space group at high temperature and undergoes a structural transition to $P2_1/c$ at 310 K and then to an unknown structure below 127 K, followed by an antiferromagnetic transition at 24 K.[24] In contrast, Kornyakov reported a four-time large unit cell with $P\bar{3}$ space group at 296 K.[20] By substituting 20% Cu with Zn to form CsClCu$_4$ZnV$_2$O$_{10}$, polycrystalline powders did not show conventional magnetic order or spin glass behavior down to 1.8 K.[24,26] Theoretical calculations predicted that the further doped averievite CsClCu$_3$Zn$_2$V$_2$O$_{10}$ is a QSL candidate.[24] Moving from V to P, it was reported that the antiferromagnetic transition $T_N$ of CsClCu$_5$P$_2$O$_{10}$ powders was suppressed to 3.8 K and the structural transition was shifted to 12 K.[21] Theoretical calculations revealed that substitution of V by P in averievite causes chemical pressure, leading to stronger interlayer coupling between Cu kagome and Cu triangle atoms and larger degree of magnetic frustration.[27] In order to address the existing key fundamental questions including the temperature dependent crystal structure, whether Zn doped averievites are QSL candidates, and the direction dependent physical properties, bulk single crystals are highly demanded; however, up to date, only sub-millimeter sized single crystals of averievite CsClCu$_5$V$_2$O$_{10}$ (0.42×0.40×0.05 mm$^3$) and CsClCu$_5$P$_2$O$_{10}$ (0.12×0.12×0.03 mm$^3$) has been reported.[20]

In this contribution, we report the successful growth of CsClCu$_5$P$_2$O$_{10}$ single crystals with dimensions of 3~5 mm on edge (**Figure 1a**) using flux method. The availability of bulk CsClCu$_5$P$_2$O$_{10}$ single crystals not only promotes structural study across first-order transitions but also provides an ideal platform for measurements of direction-dependent physical properties for deep understanding in this averievite. Combining synchrotron X-ray powder diffraction and single crystal diffraction, we report a previously unidentified first-order structural transition at 225 K, transforming from high-temperature centrosymmetric $P\bar{3}m1$ to low-temperature noncentrosymmetric $P321$ space group. Such a transition was corroborated by a pair of peaks in the differential scanning calorimetry curves and thermal hysteresis between warming and cooling in magnetic susceptibility. Strong anisotropic magnetic properties were observed for the first time in averievite. A magnetic phase diagram, consisting of a paramagnetic state, two antiferromagnetic states and four field-induced states, has been constructed based upon the magnetic susceptibility and magnetization data with magnetic fields perpendicular to the *ab* plane. Our findings suggest that the quasi-2D $CsClCu_5P_2O_{10}$ is an excellent platform for exploring novel quantum states.

## 2. Materials and Methods

**Crystal growth.** Single crystals of $CsClCu_5P_2O_{10}$ were grown using flux method for the first time in sealed quartz tubes. CuO (Macklin, AR), $Cu_2P_2O_7 \cdot H_2O$ (Macklin, 99.99%), and CsCl (Macklin, AR) were weighed with a molar ratio of 3:1:1, and then mixed with flux in a mass ratio of 1:5 where flux is $CsCl/CuCl_2$ = 7:3 (mass ratio). The $CuCl_2$ was obtained by preheating $CuCl_2 \cdot 2H_2O$ (Aladdin, AR) at 150 ºC for 12 h. The mixture was loaded into a quartz tube, sealed, then heated to 600 ºC, held at this temperature for 1 day, and then cooled to 400 ºC at a rate of 2 ºC/h, followed by furnace cooling to room temperature. Black hexagonal single crystals were obtained after removing flux using deionized water. **Figure 1a** shows a typical hexagonal black $CsClCu_5P_2O_{10}$ single crystal with dimensions of 3.12×2.86×0.97 mm³ grown from $CsCl/CuCl_2$ flux.

**Synchrotron X-ray single crystal diffraction (SXRD).** SXRD data were collected with a 1M Pilatus area detector using synchrotron radiation ($\lambda$ = 0.41328 Å) at 300, 200,100 K and 6K at Beamline 15-ID-D (NSF's ChemMatCARS) at the Advanced Photon Source, Argonne National Laboratory. Single crystals were mounted to the tip of glass fiber and measured using a Huber 3-circle diffractometer. Indexing, data reduction, and image processing were performed using Bruker APEX4 software.[28] The structure was solved by direct methods and refined with full matrix least-squares methods on $F^2$. All atoms were modeled using anisotropic ADPs, and the refinements converged for $I > 2\sigma(I)$, where $I$ is the intensity of reflections and $\sigma(I)$ is the standard deviation. Calculations were performed using the SHELXTL crystallographic software package.[28] Details of crystal parameters, data collection, and structure refinement at 300, 200,100, and 6 K are summarized in **Table S1**, selected bond lengths (Å) and angles (°) are presented in **Table S2**. Further details of the crystal structure investigations may be obtained from the joint CCDC/FIZ Karlsruhe online deposition service by quoting the deposition number CSD 2263860, 2263800, 2263806 and 2263799.

**High-resolution synchrotron X-ray powder diffraction (HRPXRD).** HRPXRD data of $CsClCu_5P_2O_{10}$ were collected in the $2\theta$ range of 0.5-50° with a step size of 0.001° and a step time of 0.1 s at 100, 200 and 300 K with an X-ray wavelength of $\lambda$=0.45903 Å at Beamline 11-BM at the Advanced Photon Source, Argonne National Laboratory. Samples were prepared by loading pulverized crystals into a Φ0.8 mm Kapton capillary, which was then installed on a magnetic sample base used by the beamline sample changer. The sample was spun continuously at 5600 rpm during data collection. An Oxford Cryostream 700 Plus $N_2$ gas blower was used to control temperature below room temperature. Diffraction patterns were recorded on warming, first at 100 K, then at 200 and 295 K. Data were analyzed with the Rietveld method using GSAS-II software [29]. Crystal structures from SXRD were used as starting models, and the refined parameters include scale, background, unit cell parameters, domain size, microstrain, atomic positions and thermal parameters. Isotropic domain size and generalized microstrain models were used.

**Differential Scanning Calorimetry (DSC).** DSC was used to check if there was any structural phase transition in $CsClCu_5P_2O_{10}$. Data between 113 and 353 K was collected on a Mettler Toledo TGA/DSC3+ with 28.03 mg $CsClCu_5P_2O_{10}$ pulverized crystals placed in an Al pan at a rate of 10 K/min on warming/cooling.

**Heat capacity ($C_P$).** The PPMS was used to measure the $C_P$ of $CsClCu_5P_2O_{10}$ with about 1.0 mg at the Synergetic Extreme Condition User Facility (SECUF), Institute of Physics, Chinese Academy of Sciences. A crystal was ground into powder, mixed with N Grease, and then transferred to a sample stage. Data was collected between 19.8 K to 1.8 K with a step size of 0.1 K.

**Magnetic Susceptibility.** DC and AC magnetic susceptibility data of $CsClCu_5P_2O_{10}$ using a single crystal of 16.4 mg were collected using the Quantum Design MPMS3 at the SECUF, Institute of Physics, Chinese Academy of Sciences. For a magnetic field perpendicular to the *ab* plane ($\mu_0 H \perp ab$), the crystal sandwiched between two quartzes was then inserted and fixed in a brass tube. ZFC-W (zero-field cooling, data collection on warming), FC-C (field cooling, data collection on cooling) and FC-W (field cooling, data collection on warming) were collected between 1.8 and 30 K with a heating rate of 3 K/min under an external magnetic field of 0.001, 0.01, 1.0, 2.0, 2.5, 3.0, 3.5, 3.65, 3.8, 4.0, 4.5, 5.0, 5.4 and 7.0 T, respectively. Magnetization data as a function of the magnetic field were collected between -7 T to +7 T with a sweep mode of 150 Oe/s at 1.8, 2.15, 2.5, 6.0, 10.0, 12.6, 13.6 and 15.0 K, respectively. For the magnetic field parallel to the *ab* plane ($\mu_0 H // ab$), the hexagonal surface of the crystal was attached to the plane of semicircular quartz columns using GE varnish. ZFC-W, FC-C and FC-W data were collected between 1.8 and 30 K, the heating rate is 3 K/min under magnetic fields of 0.001, 0.01 and 1.0 T, respectively. AC magnetic data was collected between 8.8 and 9 K with a step size of 0.02 K at 10, 97, 496 and 746 Hz, respectively. The Quantum Design PPMS Dynacool-9 was used to measure the DC magnetic susceptibility of $CsClCu_5P_2O_{10}$ single crystals between 2 and 300 K for a magnetic field perpendicular to the *ab* plane and parallel to the *ab* plane. ZFC-W, FC-C and FC-W data were collected using the sweep mode with a heating rate of 3 K/min under a magnetic field of 2000 Oe. Magnetization vs magnetic field data was collected between -9 T and +9 T using the sweep mode with a rate of 150 Oe/s at 2.0, 3.0, 5.0, 10.0, 12.0, 15.0, 20.0, 30.0 and 300 K, respectively. Measurements on other single crystals were performed to confirm the magnetic properties of the material.

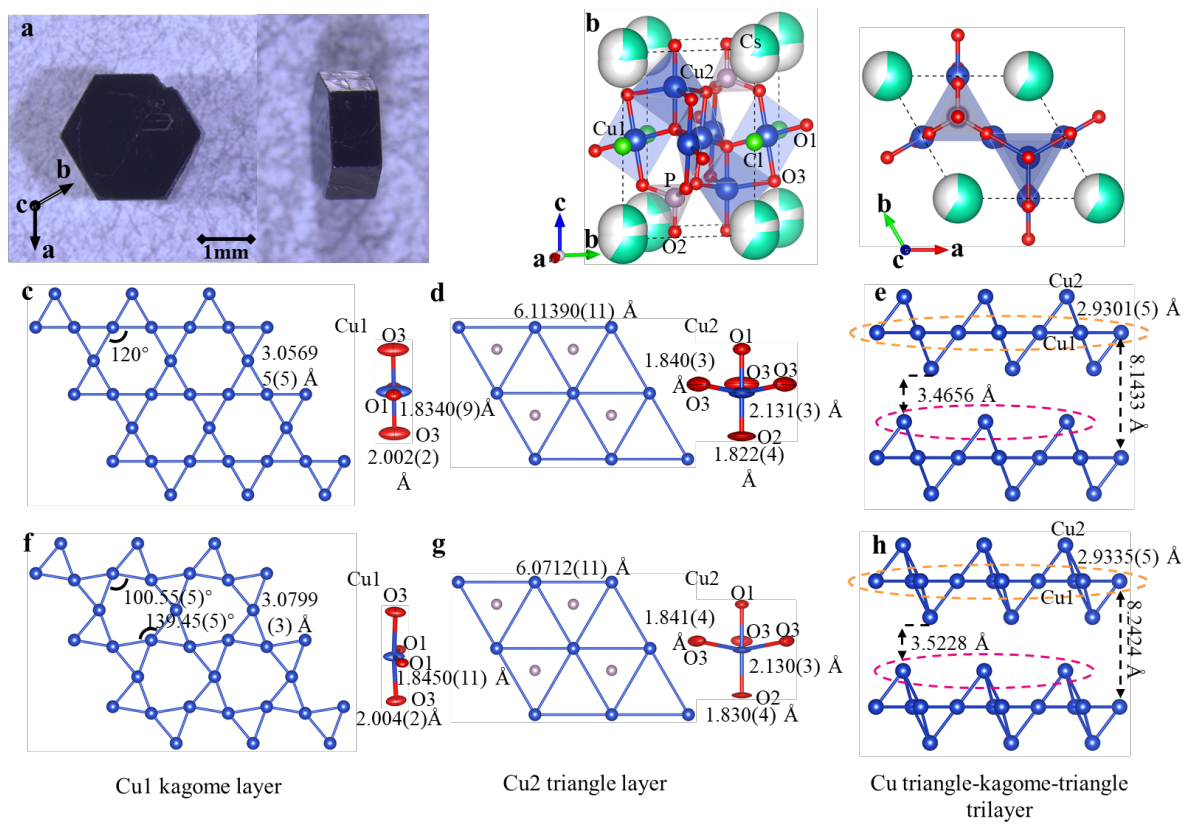

**Figure 1. Single crystals and crystal structures of $CsClCu_5P_2O_{10}$.** (a) Photos of a typical piece of as-grown single crystal, (b) Crystal structure of $CsClCu_5P_2O_{10}$ at 300 K in the ball-and-stick model, (c) Kagome layer consisting of Cu1 at 300 K and Cu1 local environment in the ellipsoid model, (d) Triangle layer consisting of Cu2 with P atoms residing in the middle of Cu triangles at 300 K, and Cu2 local environment in the ellipsoid model, (e) Triangle-kagome-triangle trilayer at 300 K, (f) Distorted kagome layer consisting of Cu1 at 100 K and Cu1 local environment in the ellipsoid model, (g) Triangle layer consisting of Cu2 at 100 K and Cu2 local environment in the ellipsoid model, (h) Triangle-kagome-triangle trilayer at 100 K.

## 3. Results

**Crystal structure.** The crystal structure of $CsClCu_5P_2O_{10}$ at various temperatures was determined using synchrotron X-ray single crystal diffraction at NSF's ChemMatCARS at Argonne National Laboratory. At room temperature, $CsClCu_5P_2O_{10}$ crystallizes in the hexagonal space group $P\bar{3}m1$ (No. 164) with lattice parameters of $a = b = 6.1139(1)$ Å and $c = 8.1433(3)$ Å, and $Z = 1$ (**Table S1**). **Figure 1b** shows the crystal structure of $CsClCu_5P_2O_{10}$ at room temperature using the ball-and-stick model. There are two Cs atoms, two Cu atoms, one Cl atom, one P atom and three O atoms in the asymmetric unit. Cs atoms are surrounded by 12 oxygen atoms and 2 Cl atoms, and Cs atoms are disordered with an occupancy of 0.59(11) for Cs1 at (0, 0, 1) and 0.21(5) for Cs2 at (0, 0, 0.946(3)). There are two types of local environments for Cu: one (Cu1) is coordinated by four O atoms with bond length of 1.8340(9)-2.002(2) Å (planar environment), and the other (Cu2) is surrounded by five O atoms with bond length in the range of 1.822(4)-2.131(3) Å (trigonal bipyramids). The P atom is coordinated by four O atoms, forming a tetrahedra with P-O bond distances in the range of 1.485(4)-1.523(3) Å (**Table S2**). The four-coordinated Cu1 form a kagome lattice in the *ab* plane with a Cu-Cu distance of 3.05695(5) Å (**Figure 1c**), and the five-coordinated Cu2 form triangles with a Cu-Cu distance of 6.11390(11) Å (**Figure 1d**). The Cu1 on each triangle is connected to Cu2 with Cu1-Cu2 = 2.9301(5) Å, forming triangle-kagome-triangle trilayers (**Figure 1e**). These trilayers stagger along the *c* direction, with 3.4656 Å between adjacent trilayers, forming a quasi-2D structure with O, Cs and Cl for charge balance.

At 100 K, the symmetry of $CsClCu_5P_2O_{10}$ is lowered and characterized by the noncentrosymmetric space group *P*321 (No. 150). The asymmetric unit contains two Cs atoms, two Cu atoms, one Cl atom, one P atom and three O atoms. Cu1 is still coordinated by four oxygen atoms; however, Cu1 and the four oxygen atoms no longer lie in the same plane (**Figure 1f**). The Cu1 atoms in the *ab* plane form a distorted kagome lattice with Cu-Cu-Cu angles of 100.55(5)° and 139.45(5)° and a distance of 3.0799(3) Å between the nearest Cu1 atoms (**Figure 1f**). The distortion increases with decreasing temperature, and angles of 97.256(14)° and 142.744(14)° are obtained at 6 K (**Figure S2**). Cu2 is surrounded by 5 oxygen atoms to form trigonal bipyramids (**Figure 1g**). Cu1-O, Cu2-O and P-O polyhedrals form a trangle-distorted kagome-trangle trilayer, and these trilayers stagger along the c direction (**Figure 1h**) to form a quasi-2D structure with Cs and Cl for charge balance.

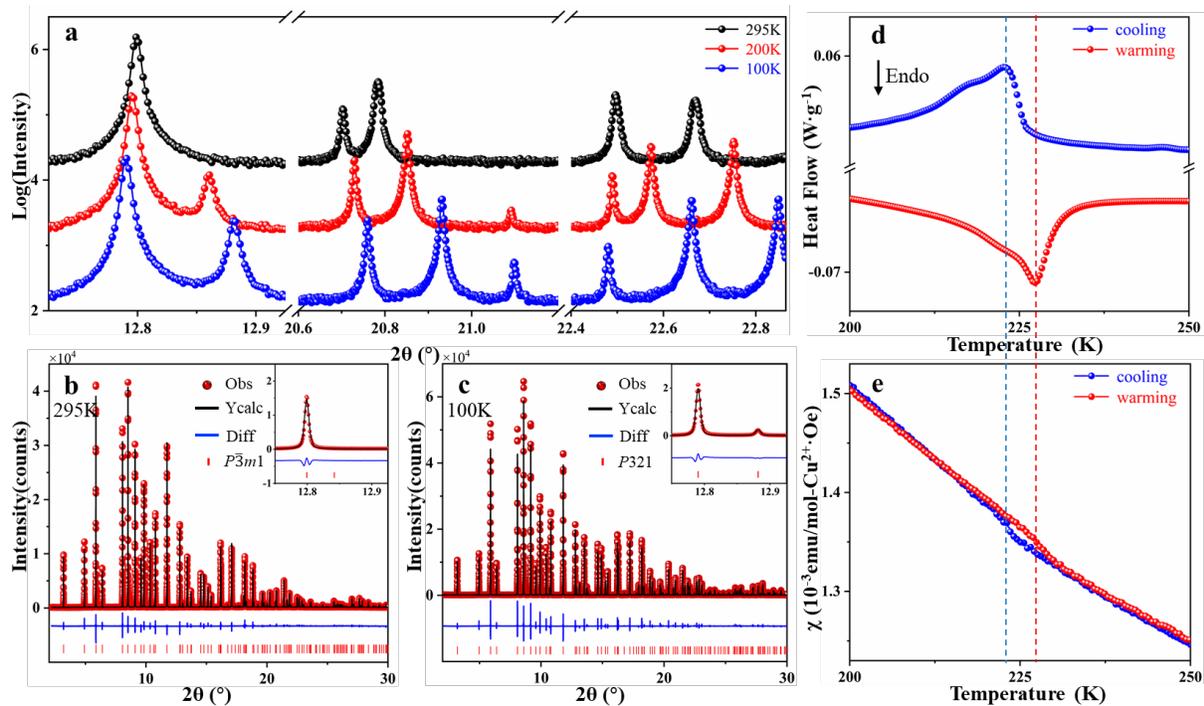

**Figure 2. Structural transition of $CsClCu_5P_2O_{10}$.** (a) High-resolution synchrotron X-ray powder diffraction pattern of $CsClCu_5P_2O_{10}$ in the $2\theta$ range of 12.725-12.925°, 20.6-21.97°, and 22.4-22.87° measured at 100, 200, and 295 K. (b) Rietveld refinement on the high-resolution synchrotron X-ray powder diffraction data of $CsClCu_5P_2O_{10}$ collected at 295 K in the $2\theta$ range of 2-30° using $P\bar{3}m1$. (c) Rietveld refinement on the high-resolution synchrotron X-ray powder diffraction data of $CsClCu_5P_2O_{10}$ collected at 100 K in the $2\theta$ range of 2-30° using *P*321. (d) DSC

data measured on cooling and warming in the temperature range of 200-250 K. (e) Magnetic susceptibility in the range of 200-250 K.

**First-order structural transition.** The structural transition between 300 and 100 K is reported here for the first time. High-resolution synchrotron X-ray powder diffraction data were collected at 11-BM at the Advanced Photon Source, Argonne National Laboratory in order to verify the existence of a structural transition. **Figure 2a** shows selected areas of the variable temperature diffraction data. Clearly, extra diffraction peaks are observed at 200 and 100 K, indicating a lower symmetry compared with 295 K. Such a structural transition occurring between 295 K and 200 K has not been reported previously, and an asymmetric peak shape is observed at 200 and 100 K, which is consistent with the previous report by Winiarski et al.[21] **Figure 2b** shows the Rietveld refinement of powder diffraction data collected at 295 K using the single crystal structural model as a starting point. The Rietveld refinement converged to $R_{wp}$ = 8.831% and GOF = 1.51 with lattice parameters of $a = b = 6.1796(1)$ Å and $c = 8.2368(1)$ Å. **Figure 2c** shows the Rietveld refinement of data at 100 K using the $P321$ space group from SXRD, and all peaks are indexed. Rietveld refinement converged to $R_{wp}$ = 11.784 % and GOF = 2.18, and the obtained lattice parameters are $a = b = 6.1315(1)$ Å and $c = 8.2424(1)$ Å.

X-ray diffraction on single crystals and powders unambiguously demonstrates the existence of a structural phase transition in $CsClCu_5P_2O_{10}$ on cooling. To further understand at what temperature the transition occurs, we carried out DSC measurements. **Figure 2d** shows the low-temperature DSC curves between 200 and 250 K. An endothermic peak at 227 K in the heating curve and a corresponding exothermic peak at 223 K in the cooling curve were clearly observed. A pair of endo- and exothermic peaks in DSC strongly supports that the transition is of first-order in nature. This first-order transition is corroborated by the thermal hysteresis in the temperature-dependent magnetic susceptibility data shown in **Figure 2e**.

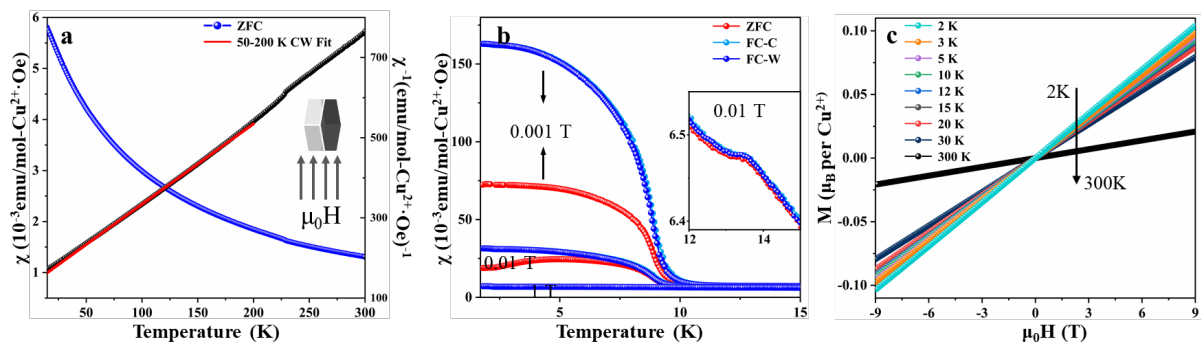

**Figure 3. Magnetic properties of $CsClCu_5P_2O_{10}$ with applied magnetic field parallel to the *ab* plane.** (a) Magnetic susceptibility under a magnetic field of 0.2 T. (b) Temperature-dependent DC magnetic susceptibility in the temperature range of 1.8-15 K under applied magnetic fields of 0.001, 0.01, and 1 T, respectively. (c) Magnetization as a function of applied magnetic field at various temperatures (2, 3, 5, 10, 12, 15, 20, 30, and 300 K).

**Magnetic properties with magnetic field parallel to the *ab* plane. Figure 3a** shows the in-plane temperature-dependent DC magnetic susceptibility under an applied field of 0.2 T in the temperature range of 15-300 K. Curie-Weiss fit in the range of 50-200 K leads to $\mu_{eff}$ = 1.96 $\mu_B$ and $\theta_{CW} = -58$ K, and $\mu_{eff} = 1.97$ $\mu_B$ and $\theta_{CW} = -69$ K in the range of 230-300 K. The negative Weiss temperature indicates strong antiferromagnetic interactions and the obtained effective moments are consistent with the expected value of 1.73 $\mu_B$ for $Cu^{2+}$ with $S = 1/2$. **Figure 3b** shows the in-plane DC magnetic susceptibility as a function of temperature in the range of 1.8-15 K under magnetic fields of 0.001, 0.01 and 1.0 T, respectively. The feature at around 13.6 K is indicative of antiferromagnetic order, and the bifurcation between ZFC and FC at around 8.9 K suggests either ferromagnetic order or spin glass behavior. However, the magnetization as a function of the applied magnetic field at low temperature does not exhibit any hysteresis (**Figure 3c**), indicating no ferromagnetism. Preliminary AC magnetic susceptibility data (**Figure S3**) in the range of 10-746 Hz show a peak around 8.9 K, but no apparent frequency dependence is observed. Whether the anomaly at 8.9 K is spin glass requires further investigation.

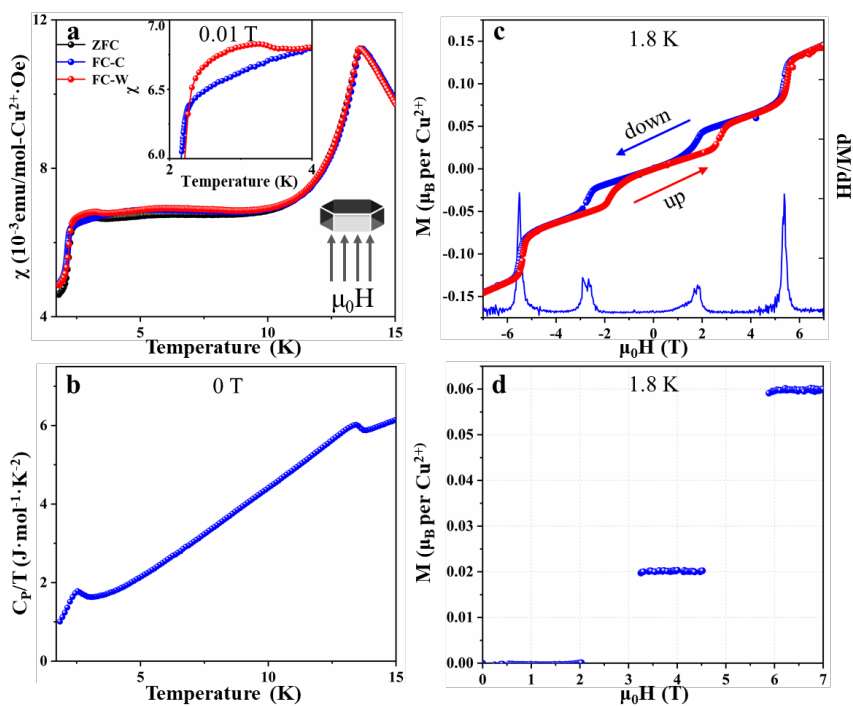

**Figure 4. Magnetic properties of CsClCu$_5$P$_2$O$_{10}$ with applied magnetic fields perpendicular to the *ab* plane.** (a) The temperature dependence of the magnetic susceptibility (ZFC-W, FC-C, FC-W) at 0.01 T. (b) Specific heat in the temperature range of 1.8-15 K. (c) Magnetization as a function of magnetic fields (blue: +7 T to -7 T, red: -7 T to +7 T) at 1.8 K and first-order derivatives of magnetization from +7 T to -7 T. (d) Net magnetization with linear components subtracted at 1.8 K, showing two magnetization plateaus.

**Magnetic properties with magnetic field perpendicular to the *ab* plane.** **Figure S4a** shows the out-of-plane magnetic susceptibility of CsClCu$_5$P$_2$O$_{10}$ as a function of temperature under an external magnetic field of 0.2 T. Above 15 K, a slope change and thermal hysteresis at around 225 K are observed (**Figure 2e**). We performed Curie-Weiss fit (**Figure S4b**) for temperature above and below the structural transition at around 225 K: (1) in the range of 50-200 K the fit leads to $\mu_{eff} = 1.77$ $\mu_B$ and $\theta_{CW} = -62$ K, and (2) in the range of 230-300 K, the fit results in $\mu_{eff} = 1.92$ $\mu_B$ and $\theta_{CW} = -120$ K. The calculated effective moments are all close to the expected value of 1.73 $\mu_B$ for $Cu^{2+}$ with $S = 1/2$. The negative Weiss temperatures indicate strong antiferromagnetic interactions. Note that across the structural transition at ~225 K, the change of Weiss temperature is remarkable, suggesting that the structural change strongly affects the interaction of the magnetic moments out of the *ab* plane.

Now we move to the magnetic susceptibility below 15 K. **Figure 4a** shows the temperature dependence of the magnetic susceptibility (ZFC-W, FC-C, FC-W) under 0.01 T in the temperature range of 1.8-15 K (see **Figure S4c** for magnetic susceptibility under other fields), and **Figure S4d** presents the first-order derivatives of the corresponding magnetic susceptibility. Three anomalies (13.6, 3.71 and 2.18 K) are clearly observed under 0.001 T. The anomaly at 13.6 K under a magnetic field of 0.001 T is suppressed to 9.55 K under 3 T and then to below 1.8 K under 7 T. The strong response to the magnetic field indicates that this anomaly is of long-range antiferromagnetic order (labeled $T_{N1}$). **Figure 4a inset** shows the temperature dependence of the magnetic susceptibility on warming and cooling in the temperature range of 1.8-5 K under 0.01 T (for other magnetic fields, see **Figure S5**). A hysteresis between 3.71 and 2.28 K is clearly observed, implying a first-order phase transition. The loop changed to 1.81-5.49 K under 3 T and then to below 1.8 K under 4.5 T. For the third anomaly at 2.18 K, it also exhibits response to the magnetic field (2.18 K under 0.001 T to below 1.8 K under 2.5 T), indicating another antiferromagnetic transition (labeled $T_{N2}$). **Figure 4b** shows the heat capacity data of $CsClCu_5P_2O_{10}$ in the temperature range of 1.8-15 K under zero magnetic field. Two anomalies (13.6 and 2.15 K) are observed, consistent with the magnetic susceptibility data shown in **Figure 4a**.

**Magnetization plateau.** **Figure 4c** shows the out-of-plane magnetization as a function of the magnetic field at 1.8 K, and the inset shows the first-order derivative of the magnetization curve (see **Figure S6** for MH curves at other temperatures). At temperatures higher than 15 K, the magnetization is linear and no hysteresis is observed, consistent with a paramagnetic state. As the temperature decreases, the magnetization deviates from linear behavior. The field-dependent magnetization exhibits a single step between 13.6 and 5 K, and two steps below 5 K. By subtracting the linear component, two magnetization plateaus were obtained (**see Figures 4d and S7-S9**). Specifically, the magnetization at 1.8 K consists of five parts in the first quadrant: three linear components and two steps. To obtain the magnetization plateau, we first fit the linear parts using $M = a\mu_0 H + b$, and then subtract $a\mu_0 H$ from the original data. Interestingly, the magnetization M = 0.02 $\mu_B$ per $Cu^{2+}$ between 3.2-4.7 T (plateau 1) is exactly one third of that M = 0.06 $\mu_B$ per $Cu^{2+}$ between 5.9-7.0 T (plateau 2). The intercept of the fitted curve on the *y*-axis in **Figure S6a** corresponds to the magnetization plateaus. The estimated saturated moment of per $Cu^{2+}$ is $M_S = gS\mu_B = 2\times1/2\mu_B = 1.0$ $\mu_B$. The two plateaus M=0.02 and 0.06 $\mu_B$ per $Cu^{2+}$ are only 1/50 and 3/50 of the saturated $M_S$ (see **Figure 4d**), suggesting the existence of two quantum magnetic states.

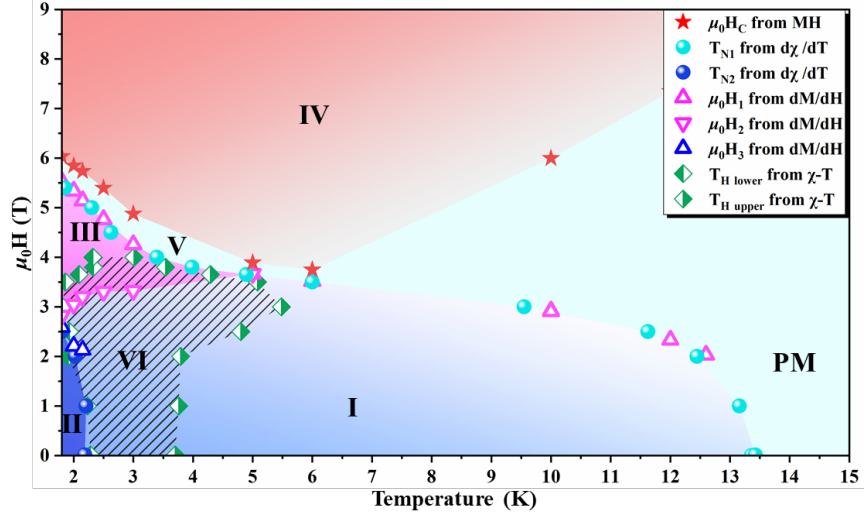

**Figure 5.** Magnetic phase diagrams of $CsClCu_5P_2O_{10}$ with $\mu_0H \perp ab$.

**Magnetic phase diagram. Figure 5** shows the magnetic phase diagram of $CsClCu_5P_2O_{10}$ with a magnetic field perpendicular to the *ab* plane, summarizing the magnetic susceptibility and magnetization data. As can be seen, the phase diagram consists of seven states. $T_{N1}$, the peak of the first-order derivative of $\chi$, delineates the boundary between the long-range antiferromagnetically ordered (State I) and paramagnetic states of the system, and $T_{N2}$ is the dividing line between State I (AFM1) and II (AFM2). The upper and lower limits of State III are determined by the positions of the peaks in the first-order derivatives of the MH curves, $\mu_0H_1$ and $\mu_0H_2$. Its upper limit coincides with $T_{N1}$. The lower limit of State IV (probably a spin polarized state) is determined by the critical magnetic field ($\mu_0H_C$) of plateau 2 in the MH curve (**Figure S6**). State V, corresponding to the part of the MH curve between $\mu_0H_1$ and $\mu_0H_C$, is located between III and IV and it is an intermediate state between them. State VI is determined by the thermal hysteresis between FC-C and FC-W (**Figure S5**), indicating a first-order phase transition. The left boundary of VI coincides with the right-side boundary of II. There exist overlaps between State VI and III as well as State VI and I, implying that there is a correlation between the structural transition and AFM1 and the magnetization plateau (III). For magnetic structures of $CsClCu_5P_2O_{10}$ under various magnetic fields, single crystal and/or powder neutron diffraction data are essential to elucidate the phase diagram.

It is well known that the averievite compounds show structural and magnetic transitions on cooling,[21,24,26,30] for example, two structural transitions and one antiferromagnetic transition were found in polycrystalline V-averievite,[24] and double antiferromagnetic transitions were reported in polycrystalline $RbClCu_5P_2O_{10}$.[30] However, no study on anisotropic properties has been found. Here we succeed in growing bulk $CsClCu_5P_2O_{10}$ single crystals, and report for the first time the large anisotropic magnetic properties and field-induced quantum magnetic states under external magnetic fields. This is related to the triangle-kagome-triangle trilayer, which is a common feature of the averievite family, and the smaller ion size of P.[30] It has been well established that magnetization plateau is the result of the interaction of competing phases in the system from a disorder to an order state.[31,32] For example, 1/6, 1/3, 2/3, and 5/6 plateaus were observed in kagome-lattice HoAgGe.[33] Taking the 1/3 plateau in the triangular lattice with $S = 1/2$ for example, the magnetic moment is partially polarized under magnetic field, forming an "up-up-down" spin structure, so the macroscopic magnetization is expressed as 1/3 of the saturation magnetic strength.[34] For $CsClCu_5P_2O_{10}$, the magnetization plateau is only 1/50 and 3/50 of the saturated moment of $Cu^{2+}$ ($S$ = 1/2), suggesting a different mechanism compared with other common magnetization plateaus. It is intriguing to explore magnetic plateaus in $CsClCu_5P_2O_{10}$ under high magnetic fields as well as in other averievite compounds.

## 4. Discussion

In summary, we report for the first time the cascade of structural and field-induced transitions in the quasi-2D averievite $CsClCu_5P_2O_{10}$ single crystals. A previously unreported structural transition from centrosymmetric $P\bar{3}m1$ to noncentrosymmetric $P321$ was determined by combining synchrotron X-ray single crystal and powder diffraction, differential scanning calorimetry and magnetic properties. The noncentrosymmetric crystal structure ($P321$ space group) below the first-order transition in single crystals suggests that the averievite material may have potential applications in the fields of nonlinear optics and piezoelectricity. Large anisotropy was observed in the magnetic properties of $CsClCu_5P_2O_{10}$ due to the availability of bulk single crystals. A magnetic phase diagram, consisting of a paramagnetic state, two antiferromagnetic states and four field-induced states including two magnetization plateaus, has been constructed based upon the magnetic susceptibility and magnetization data with a magnetic field perpendicular to the $ab$ plane. Determination of the magnetic structures in the phase diagram will require other techniques such as neutron diffraction. $CsClCu_5P_2O_{10}$ is the first member of the averievite family showing large anisotropic magnetic properties and rich magnetic phase diagram under external magnetic fields. It is of great interest to investigate other members of the averievite family, $(MX)_nCu_5T_2O_{10}$ (M=K, Rb, Cs, Cu; X=Cl, Br, I; n=1; T=P, V). Although $CsClCu_5P_2O_{10}$ is far from the long-sought quantum spin liquid state, substution of copper in the triangle layers using non-magnetic ions such as Zn and Mg provide an effective strategy for exploring novel quantum states including quantum spin liquid.

## ASSOCIATED CONTENT

**Supporting Information**.

Supplementary data to this article can be found online at*** The X-ray diffraction pattern of $CsClCu_5P_2O_{10}$ crystal; Kagome layer consisting of Cu1 at various temperatures; In-plane AC magnetic susceptibility at various frequencies; Magnetic properties of $CsClCu_5P_2O_{10}$ with applied magnetic fields perpendicular to the $ab$ plane; Temperature dependent magnetic susceptibility under various magnetic fields; Magnetization is a function of magnetic fields at various temperatures with applied magnetic fields perpendicular to the $ab$ plane; Magnetization plateau at 1.8 K ($\mu_0H \perp ab$); Net magnetization with linear components subtracted for various temperatures (MPMS3 data and PPMS DynaCool-9T data); Crystallographic data and refinement parameters for $CsClCu_5P_2O_{10}$ at various temperatures; Selected bond length and angles for $CsClCu_5P_2O_{10}$ at various temperatures.

## AUTHOR INFORMATION

**Corresponding Author**

**Junjie Zhang -** State Key Laboratory of Crystal Materials and Institute of Crystal Materials, Shandong University, Jinan, Shandong 250100, China; orcid.org/0000-0002-5561-1330, Email: junjie@sdu.edu.cn

**Authors**

**Chao Liu** – State Key Laboratory of Crystal Materials and Institute of Crystal Materials, Shandong University, Jinan, Shandong 250100, China

**Tieyan Chang** – NSF's ChemMatCARS, The University of Chicago, Argonne, Il 60439, USA

**Shilei Wang** – State Key Laboratory of Crystal Materials and Institute of Crystal Materials, Shandong University, Jinan, Shandong 250100, China

**Shun Zhou** – State Key Laboratory of Crystal Materials and Institute of Crystal Materials, Shandong University, Jinan, Shandong 250100, China

**Xiaoli Wang** – State Key Laboratory of Crystal Materials and Institute of Crystal Materials, Shandong University, Jinan, Shandong 250100, China

**Chuanyan Fan** – State Key Laboratory of Crystal Materials and Institute of Crystal Materials, Shandong University, Jinan, Shandong 250100, China

**Lu Han** – State Key Laboratory of Crystal Materials and Institute of Crystal Materials, Shandong University, Jinan, Shandong 250100, China

**Feiyu Li** – State Key Laboratory of Crystal Materials and Institute of Crystal Materials, Shandong University, Jinan, Shandong 250100, China

**Huifen Ren** – Synergetic Extreme Condition User Facility, Institute of Physics CAS, Beijing 101400, China

**Shanpeng Wang** – State Key Laboratory of Crystal Materials and Institute of Crystal Materials, Shandong University, Jinan, Shandong 250100, China

**Yu-Sheng Chen** – NSF's ChemMatCARS, The University of Chicago, Argonne, Il 60439, USA

**Author Contributions**

J.Z. conceived and lead the project. C.L. grew single crystals and performed in-house powder and single-crystal X-ray diffraction experiments with the help from S.Z., X.W., C.F., L.H. and F.L. C.L. carried out magnetic measurements with the help of S.L.W., S.P.W., H. R. and J.Z. T.C. and Y.C. carried out synchrotron X-ray single crystal and powder diffraction measurements. H.R. carried out heat capacity measurements. C.L. and J.Z. analyzed data. C.L. and J.Z. wrote the manuscript with contributions from all coauthors. All authors have given approval to the final version of the manuscript.

**Notes**

The authors declare no competing interest.

ACKNOWLEDGMENT
J.Z thanks Prof. Xutang Tao for providing valuable support and fruitful discussions. J.Z thanks Mr. Mingqi Zhang and Ms. Mengyao Li from Mettler Toledo for their help with low temperature DSC measurements. Work at Shandong University was supported by the National Natural Science Foundation of China (12374457 and 12074219), the TaiShan Scholars Project of Shandong Province (tsqn201909031), the QiLu Young Scholars Program of Shandong University, the Crystalline Materials and Industrialization Joint Innovation Laboratory of Shandong University and Shandong Institutes of Industrial Technology (Z1250020003), and the Project for Scientific Research Innovation Team of Young Scholars in Colleges and Universities of Shandong Province (2021KJ093). A portion of this work was carried out at the Synergetic Extreme Condition User Facility (SECUF). NSF's ChemMatCARS, Sector 15 at the Advanced Photon Source (APS), Argonne National Laboratory (ANL) is supported by the Divisions of Chemistry (CHE) and Materials Research (DMR), National Science Foundation, under grant number NSF/CHE-1834750. This research used resources of the Advanced Photon Source; a U.S. Department of Energy (DOE) Office of Science user facility operated for the DOE Office of Science by Argonne National Laboratory under Contract No. DE-AC02-06CH11357.